\documentclass[a4paper,11pt]{article}
\usepackage{jheppub}
\usepackage[T1]{fontenc}
\usepackage{subfigure}
\usepackage{overpic}
\usepackage{multirow}
\usepackage{lineno}

\title{\boldmath Measurement of $e^{+}e^{-}\to \omega\eta^{\prime}$ cross sections at $\sqrt{s}=$ 2.000 to 3.080 GeV}

\collaborationImg{\includegraphics[height=3cm,angle=90]{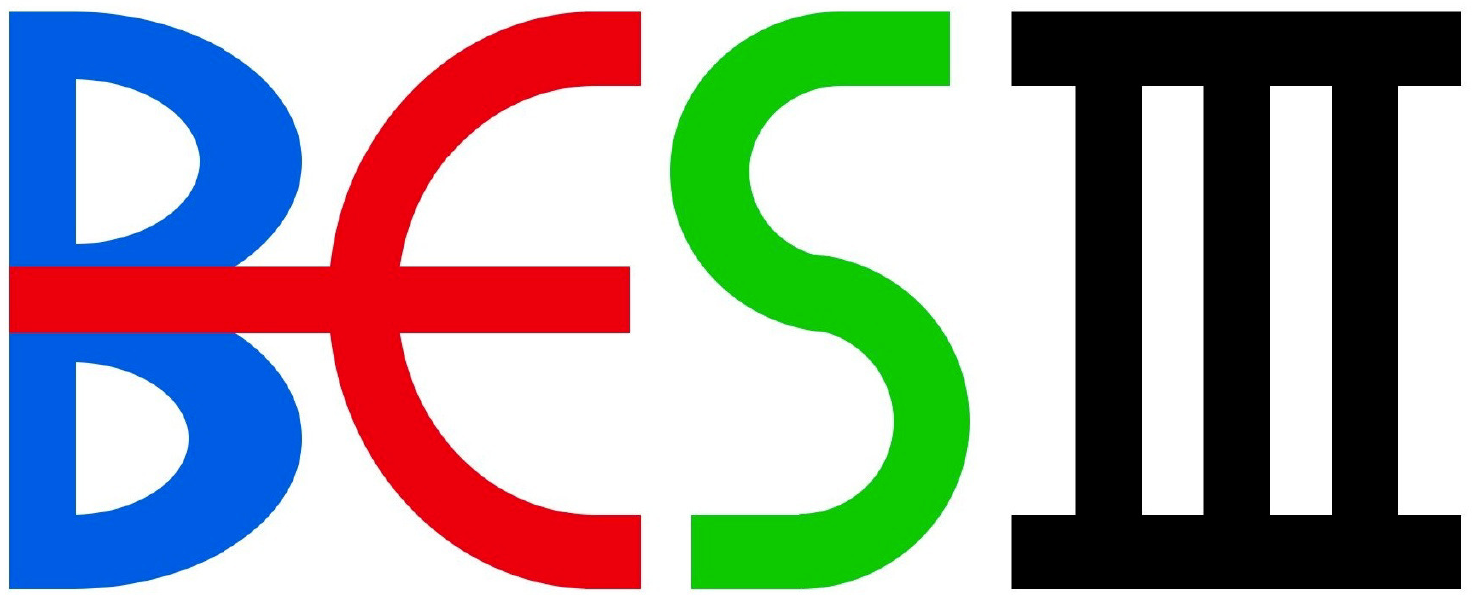}}
\collaboration{The BESIII Collaboration}

\emailAdd{besiii-publications@ihep.ac.cn}

\abstract{The Born cross sections for the process $e^{+}e^{-}\to
  \omega\eta^{\prime}$ are measured at 22 center-of-mass energies from
  2.000 to 3.080 GeV using data collected with the BESIII detector at the BEPCII collider. A resonant structure is observed with a statistical significance of 9.6$\sigma$. A Breit-Wigner fit determines its mass to be $M_R=(2153\pm30\pm31)~{\rm{MeV}}/c^{2}$ and its width to be $\Gamma_{R}=(167\pm77\pm7)~\rm{MeV}$, where the first uncertainties are statistical and the second are systematic.}

  
\arxivnumber{....}

\begin{document}                                                                                                                
    \maketitle
    \flushbottom
    \section{Introduction}
    Hadron spectroscopy offers valuable insights into the non-perturbative dynamics of the strong interaction.  According to the Particle Data Group~(PDG)~\cite{PDG}, there are several vector states of the $\omega$ family, \textit{i.e.} $J^{PC}=1^{--}$ isoscalar states, with masses around 2.2 GeV, $\omega(2205)$, $\omega(2290)$ and $\omega(2330)$, which need to be confirmed. It is noteworthy that the measured widths are usually at least twice as wide as the theoretical calculations~\cite{omega2205,omega2290,omega2330,regge_pang,jiecheng,pang2022}.
    
    The BaBar Collaboration has studied the process $e^{+}e^{-}\to\omega\pi^{+}\pi^{-}$ using the initial state radiation (ISR) method. The cross section line shape shows evidence for an isoscalar resonance structure near 2.25 GeV with a significance of 2.6$\sigma$~\cite{babar_omegapipi,babar_fit}. The BESIII Collaboration has studied the processes $e^{+}e^{-}\to\omega\eta$~\cite{bes_omegaeta}, $e^{+}e^{-}\to\omega\pi^{0}\pi^{0}$~\cite{bes_omegapi0pi0} and $e^{+}e^{-}\to\omega\pi^{+}\pi^{-}$~\cite{bes_omegapippim} at center-of-mass~(c.m.) energies~($\sqrt{s}$) from 2.000 to 3.080 GeV. Two structures around \mbox{2.2 GeV} are observed with resonance parameters \mbox{$M_{1}=2176\pm24$ ${\rm{MeV}}/c^{2}$}, \mbox{$\Gamma_{1}=89\pm50$ $\rm{MeV}$} and $M_{2}=2232\pm19$ ${\rm{MeV}}/c^{2}$, $\Gamma_{2}=93\pm53$ $\rm{MeV}$, respectively. The authors in reference~\cite{qinsong} introduced the $\omega(4S)$ and $\omega(3D)$ to represent the behavior of these enhancements in the cross section, and claimed that these two structures are caused by the interference of $\omega(4S)$ and $\omega(3D)$. To establish higher excited $\omega$ states, more theoretical and experimental efforts are desirable.

    In this paper, we present a measurement of the Born cross sections for the process $e^{+}e^{-}\to\omega\eta^{\prime}$ at $\sqrt{s}$ from 2.000 to 3.080 GeV based on 22 data samples corresponding to an integrated luminosity of 650 $\rm{pb}^{-1}$ collected by the BESIII experiment.

    \section{BESIII detector and Monte Carlo simulation}
    The BESIII detector~\cite{besiii} records symmetric $e^{+}e^{-}$ collisions provided by the BEPCII storage ring~\cite{bepcii} in the range of $\sqrt{s}$ from 2.0 to 4.95 GeV, with a peak luminosity of $1\times10^{33}~{\rm{cm}}^{-2}{\rm{s}}^{-1}$ achieved at $\sqrt{s}=3.77$ GeV. BESIII has collected large data samples in this energy region~\cite{BESIIIDATA}. The cylindrical core of the BESIII detector covers 93\% of the full solid angle and consists of a helium-based multilayer drift chamber~(MDC), a plastic scintillator time-of-flight system~(TOF), and a CsI(Tl) electromagnetic calorimeter~(EMC), which are all enclosed in a superconducting solenoidal magnet providing a 1.0 T~(0.9 T in 2012) magnetic field. The solenoid is supported by an octagonal flux-return yoke with resistive plate counter muon identification modules interleaved with steel. The charged-particle momentum resolution at 1 GeV$/c$ is 0.5\%, and the specific ionization energy loss (d$E$/d$x$) resolution is 6\% for electrons from Bhabha scattering. The EMC measures photon energies with a resolution of 2.5\%~(5\%) at 1 GeV in the barrel~(end cap) region. The time resolution in the TOF barrel region is 68 ps, while that in the end cap region is 110 ps.

    Monte Carlo (MC) simulated data samples produced with a {\sc{GEANT4}}-based~\cite{bes_g4} software package, which includes the geometric description~\cite{detvis} of the BESIII detector and the detector response, are used to optimize the event selection criteria, estimate background processes, and determine the detection efficiency. The signal MC samples for the process $e^{+}e^{-}\to\omega\eta^{\prime}$ are generated by {\sc{ConExc}}~\cite{bes_conexc} so that the contribution of the P-wave amplitude between the vector and pseudoscalar mesons in the final state is taken into account. The decay of the $\eta^{\prime}$ and $\omega$ are described by the observed amplitude patterns in $\eta^{\prime}\to\gamma\pi^{+}\pi^{-}$ and $\omega\to\pi^{+}\pi^{-}\pi^{0}$, respectively~\cite{besiii_etapDA,besiii_omegaDA}. For background studies, MC samples of inclusive hadronic events are generated with a hybrid generator that integrates {\sc{ConExc}}~\cite{bes_conexc}, {\sc{LUARLW}}~\cite{bes_luarlw} and {\sc{PHOKHARA}~\cite{bes_phokhara}}. Exclusive MC samples of $e^{+}e^{-}\to\omega\pi^{+}\pi^{-}$ and $e^{+}e^{-}\to2(\pi^{+}\pi^{-}\pi^{0})$ are generated by {\sc{ConExc}}~\cite{bes_conexc} according to published results of BESIII and BaBar~\cite{bes_omegapippim,babar_6pi}.
    
    \section{Event selection and background analysis}
    The signal process $e^{+}e^{-}\to \omega\eta^{\prime}$ is reconstructed using the decays $\omega\to \pi^{+}\pi^{-}\pi^{0}$, $\pi^{0}\to \gamma\gamma$ and \mbox{$\eta^{\prime}\to \gamma\pi^{+}\pi^{-}$}. The signal candidates are required to contain four charged pions with zero net charge and at least three photon candidates. 

    Charged tracks detected in the MDC are required to be within a polar angle~($\theta$) range of $\vert\!\cos\theta\vert<0.93$, where $\theta$ is defined with respect to the $z$ axis, defined as the symmetry axis of the MDC. For each charged track, the distance of closest approach to the interaction point~(IP) is required to be within 10 cm in the beam direction and within 1 cm in the plane perpendicular to the beam direction. Particle identification~(PID) for charged tracks combines measurements of the d$E$/d$x$ in the MDC and the flight time in the TOF to form likelihoods $\mathcal{L}(h)(h=p,K,\pi)$ for each hadron $h$ hypothesis. Charged tracks are identified as pions when the pion hypothesis has the greatest likelihood~[$\mathcal{L}(\pi)>\mathcal{L}(K)$ and $\mathcal{L}(\pi)>\mathcal{L}(p)$]. 

    Photon candidates are identified using isolated showers in the EMC. The deposited energy of each shower must be more than 25 MeV in the barrel region~($\vert\!\cos\theta\vert<0.80$) and more than 50 MeV in the end cap region~($0.86<\vert\!\cos\theta\vert<0.92$). To exclude showers that originate from charged tracks, the angle subtended by the EMC shower and the position of the closest charged track at the EMC must be greater than $10^\circ$ as measured from IP. To suppress electronic noise and showers unrelated to the event, the difference between the EMC time and the event start time of the photon candidate is required to be within \mbox{[0,700] ns}.

\begin{figure}[t]
        \centering
        \begin{overpic}[width=0.48\textwidth]{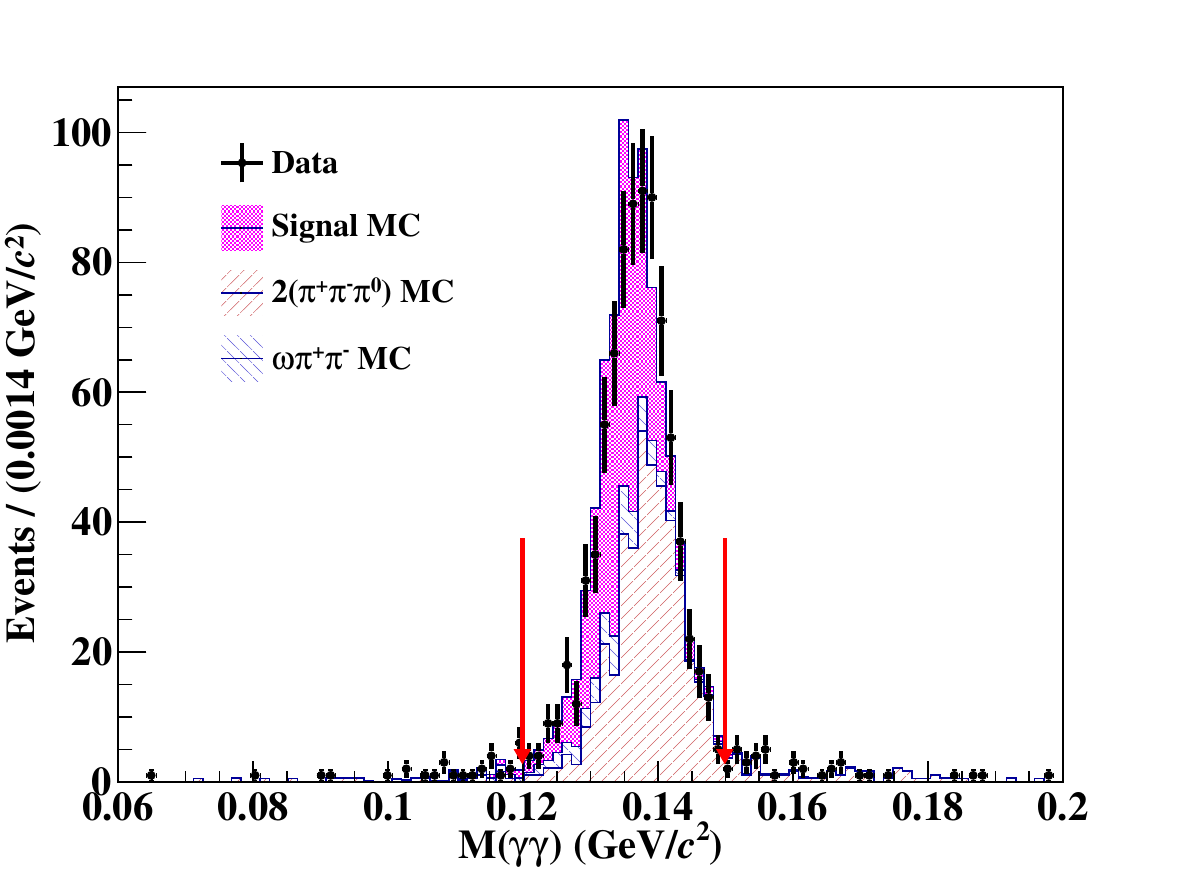}
        \put(75,60){(a)}
        \end{overpic}
        \begin{overpic}[width=0.48\textwidth]{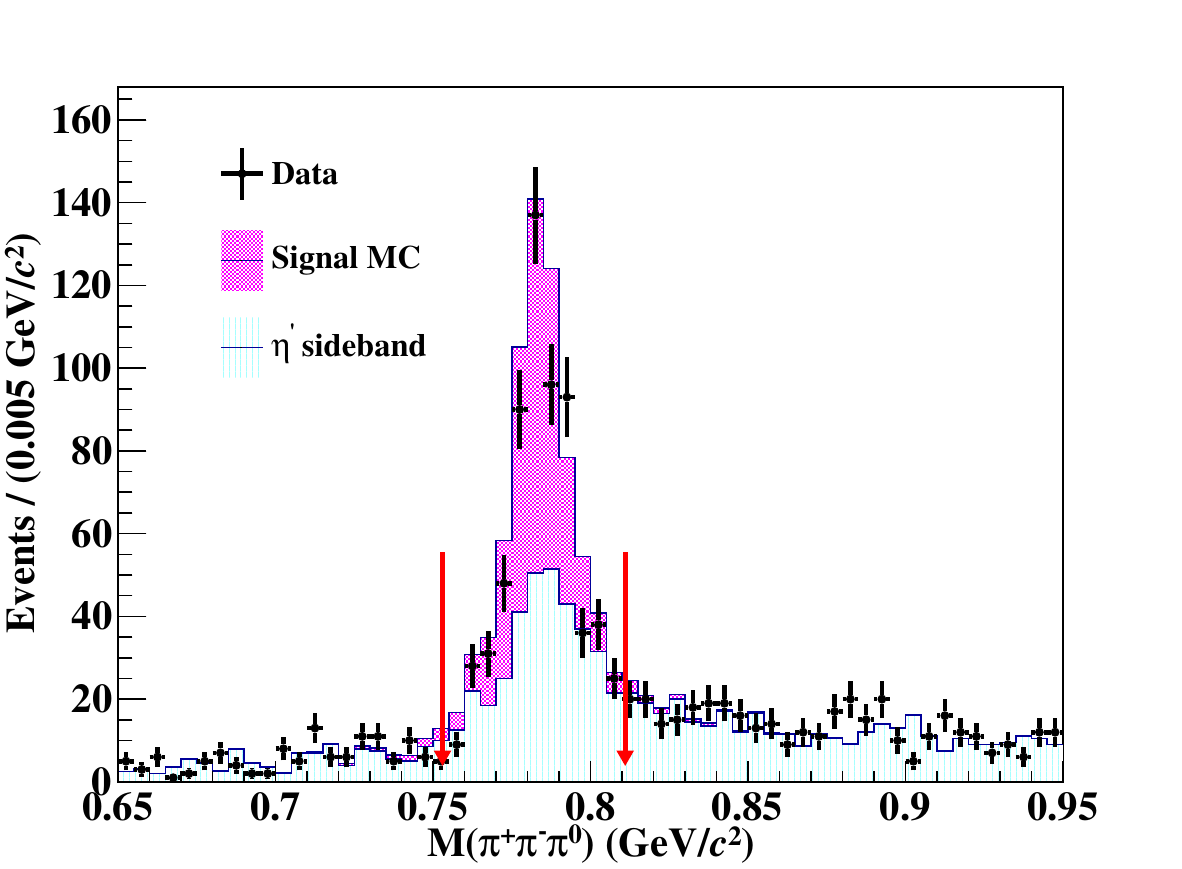}
        \put(75,60){(b)}
        \end{overpic}
        \begin{overpic}[width=0.48\textwidth]{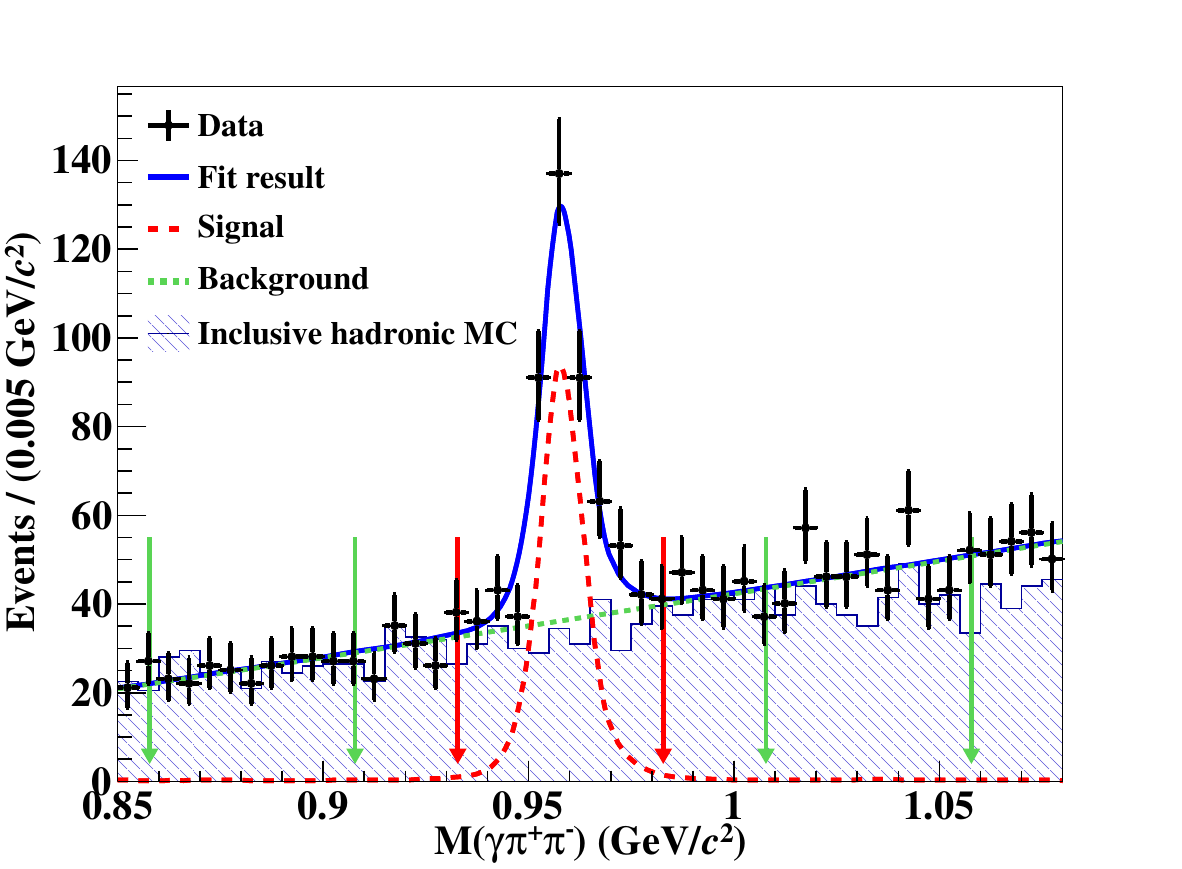}
        \put(75,60){(c)}
        \end{overpic}
        \flushleft
        \caption{(a) Distribution of $M(\gamma\gamma)$, where the (black) dots with error bars are data, and the shaded histogram are stacked MC samples of signal MC, $2(\pi^{+}\pi^{-}\pi^{0})$ MC and $\omega\pi^{+}\pi^{-}$ MC. (b) Distribution of $M(\pi^{+}\pi^{-}\pi^{0})$ in the $\eta^{\prime}$ signal region, where the (black) dots with error bars are data, and the shaded histogram are the stacked signal MC sample and non-$\eta^{\prime}$ events estimated by the $\eta^{\prime}$ sideband. (c) Fit to the $M(\gamma\pi^{+}\pi^{-})$ distribution for events in the $\pi^0$ and $\omega$ signal regions, where the (black) dots with error bars are data, the (blue) solid curve is the total fit result, the (green) dashed curve indicates the fitted background shape, and the (red) dash-dotted curve is the fitted signal shape. The vertical lines indicate the signal~(red) and sideband regions~(green).  All distributions shown are from the data sample at $\sqrt{s} = 2.125$~GeV.}
        \label{fig:2125signal}
    \end{figure}

    To suppress the background and improve the resolution of kinematic quantities, a four-constraint~(4C) kinematic fit imposing energy-momentum conservation is carried out under the hypothesis of $e^{+}e^{-}\to 2(\pi^{+}\pi^{-})3\gamma$. If there are more than three photon candidates in one event, the combination with the minimum $\chi_{{\rm{4C}}}^{2}$ is retained for further analysis. The candidate events are required to satisfy $\chi_{{\rm{4C}}}^{2}<60$. To suppress the contamination from the processes \mbox{$e^{+}e^{-}\to 2(\pi^{+}\pi^{-})\pi^{0}$} or $e^{+}e^{-}\to 2(\pi^{+}\pi^{-}\pi^{0})$, two additional 4C kinematic fits under the hypotheses of $e^{+}e^{-}\to 2(\pi^{+}\pi^{-}\gamma)$ and $e^{+}e^{-}\to 2(\pi^{+}\pi^{-}2\gamma)$ are independently performed. Only those events which satisfy $\chi^{2}_{{\rm{4C}}}<\chi^{2}_{{\rm{4C}}}(2(\pi^{+}\pi^{-}\gamma))$ and $\chi^{2}_{{\rm{4C}}}<\chi^{2}_{{\rm{4C}}}(2(\pi^{+}\pi^{-}2\gamma))$ are retained. The photon combination with the smallest value of $|M(\gamma\gamma)-M_{\pi^{0}}|$ is used to form the $\pi^{0}$ candidate, where $M_{\pi^{0}}$ is the known $\pi^{0}$ mass~\cite{PDG}. The distribution of the $\gamma\gamma$ invariant mass, $M(\gamma\gamma)$, at \mbox{$\sqrt{s}=2.125$ GeV} is shown in Fig.~\ref{fig:2125signal}(a). The $\omega$ candidate is taken to be the $\pi^{+}\pi^{-}\pi^{0}$ combination with the smallest value of $|M(\pi^{+}\pi^{-}\pi^{0})-M_{\omega}|$, where $M_{\omega}$ is the nominal $\omega$ mass~\cite{PDG}. The distribution of the $\pi^{+}\pi^{-}\pi^{0}$ invariant mass, $M(\pi^{+}\pi^{-}\pi^{0})$, at \mbox{$\sqrt{s}=2.125$ GeV} is shown in Fig.~\ref{fig:2125signal}(b). The distribution of the invariant mass of the remaining photon and $\pi^{+}\pi^{-}$, $M(\gamma\pi^{+}\pi^{-})$, at \mbox{$\sqrt{s}=2.125$ GeV} is shown in Fig.~\ref{fig:2125signal}(c), where an $\eta^{\prime}$ signal is visible with the \mbox{$|M(\gamma\gamma)-M_{\pi^{0}}|<0.015$ GeV/$c^{2}$} and \mbox{$|M(\pi^{+}\pi^{-}\pi^{0})-M_{\omega}|<0.029$ GeV/$c^{2}$} requirements, corresponding to about 3$\sigma$ in the mass resolutions. The signal region of the $\eta^{\prime}$ is defined as \mbox{$|M(\gamma\pi^{+}\pi^{-})-M_{\eta^{'}}|<0.025$ ${\rm{GeV}}/c^{2}$}, where $M_{\eta^{\prime}}$ is the nominal $\eta^{\prime}$ mass~\cite{PDG}. The sideband region of the $\eta^{\prime}$ is defined as $0.05<|M(\gamma\pi^{+}\pi^{-})-M_{\eta^{'}}|<0.10$ ${\rm{GeV}}/c^{2}$. 

    Potential background sources are studied by analyzing inclusive $e^{+}e^{-}\to$ hadrons and exclusive MC samples. Exclusive MC samples for $e^{+}e^{-}\to\omega\pi^{+}\pi^{-}$ and $e^{+}e^{-}\to 2(\pi^{+}\pi^{-}\pi^{0})$ are generated by {\sc{ConExc}}~\cite{bes_conexc} with an amplitude model introduced in ref.~\cite{bes_omegapippim} for the process $e^{+}e^{-}\to\omega\pi^{+}\pi^{-}$ and a phase-space model for the process $e^{+}e^{-}\to 2(\pi^{+}\pi^{-}\pi^{0})$ taking into account the cross sections measured by BaBar~\cite{babar_6pi}. Simulated events are subject to the same selection procedure as the signal process. Those non-$\eta^{\prime}$ events, {\it e.g.}, $e^{+}e^{-}\to\omega\pi^{+}\pi^{-}$, form a peaking background in the $M(\pi^{+}\pi^{-}\pi^{0})$ distribution, as shown in figure~\ref{fig:2125signal}(b) and are estimated using the $\eta^{\prime}$ sideband. The background in the $M(\gamma\pi^{+}\pi^{-})$ distribution, determined from exclusive MC samples and the inclusive hadronic MC sample, is flat. Therefore, the number of signal is determined by fitting the $M(\gamma\pi^{+}\pi^{-})$ spectrum. The dominant background stems from the $e^{+}e^{-}\to 2(\pi^{+}\pi^{-}\pi^{0})$ process. The background distribution obtained from the inclusive hadronic MC sample, which has been normalized to the integrated experimental luminosity, is shown in figure~\ref{fig:2125signal}(c).

    The signal yields of the $e^{+}e^{-}\to \omega\eta^{\prime}$ process are obtained by performing an unbinned maximum likelihood fit to the $M(\gamma\pi^{+}\pi^{-})$ spectrum. The signal is described by the signal MC-simulated shape convolved with a Gaussian function that describes the difference between data and MC simulation. The background function is parametrized by a second-order Chebychev function. The corresponding fit result for data taken at $\sqrt{s}=2.125$ GeV is shown in Fig.~\ref{fig:2125signal}(c). The same event selection criteria and fit procedure are applied for all data samples at the other c.m.\ energies. The obtained signal yields are listed in table~\ref{tab:crosssection}.
    \section{Born cross section and systematic uncertainty}
    The Born cross section for $e^{+}e^{-}\to \omega\eta^{\prime}$ is calculated by
    \begin{equation}
    \label{cs}
        \sigma(\sqrt{s})=\frac{N_{\rm{sig}}}{\mathcal{L}\cdot\epsilon\cdot\mathcal{B}\cdot\frac{1}{|1-\Pi|^{2}}\cdot(1+\delta)},
    \end{equation}
    where $N_{\rm{sig}}$ is the number of signal events, $\mathcal{L}$ is the integrated luminosity, $\epsilon$ is the selection efficiency, $\mathcal{B}$ is the product of subleading branching fractions, $\mathcal{B}=\mathcal{B}_{\omega\to\pi^{+}\pi^{-}\pi^{0}}\cdot\mathcal{B}_{\eta^{\prime}\to\gamma\pi^{+}\pi^{-}}\cdot\mathcal{B}_{\pi^{0}\to\gamma\gamma}=26.0\%$, taken from the PDG~\cite{PDG}, $\frac{1}{|1-\Pi|^{2}}$ is the vacuum polarization factor~(VP)~\cite{VP}, and $1+\delta$ is the ISR correction factor, which is obtained by a QED calculation~\cite{isr_calculate} that takes the line shape of the Born cross section into account. Both $\epsilon$ and $1+\delta$ depend on the line shape of the cross sections and are determined via an iterative procedure~\cite{jingmq,bes_kskl}. The $\epsilon$ and $1+\delta$ are calculated according to the fit curve and are taken as input for the next iteration. The procedure is repeated until the difference between the final measured Born cross sections and the last one less than 0.1\%. The statistical significance of the signal for each energy point is estimated by considering the change of likelihood $\Delta(L)$ and the number of degrees of freedom in fits that include or do not include the signal function.  For energy points where the significance of the signal is less than or close to $3\sigma$, the upper limits for the number of signal events~($N_{\rm{sig}}^{\rm{up}}$) and the cross section~($\sigma^{\rm{up}}$) are obtained by performing a profile-likelihood method~\cite{upper}. The Born cross sections and upper limits at the 90\% confidence level for all 22 energy points are summarized in table~\ref{tab:crosssection}.
    \begin{table}[t]
    \centering
    \scalebox{0.7}[0.7]{
    \begin{tabular}{ccccccccc}  
  \hline
  $\sqrt{s}$ (GeV)  &$N_{\rm{sig}}$ &$N_{\rm{sig}}^{\rm{up}}$ &$\mathcal{L}$ (pb$^{-1}$) &$\epsilon\cdot(1+\delta)$ &VP &$\sigma$ (pb) &$\sigma^{\rm{up}}$(pb) &Significance ($\sigma$)\\
  \hline
       2.0000 &$10.9\pm5.6$           &<23.1  &10.1  &0.0988 &1.037  &$40.63\pm20.87\pm2.56$             &<86.13 &1.9\\ 
       2.0500 &$10.5\pm4.1$           &-      &3.3  &0.0986 &1.038  &$118.40\pm46.22\pm7.46$            &-      &3.6\\ 
       2.1000 &$23.7\pm6.8$           &-      &12.2  &0.0987 &1.039  &$72.43\pm20.78\pm4.56$             &-      &4.4\\ 
       2.1250 &$273.7\pm22.6$         &-      &108.5   &0.1000 &1.039  &$93.71\pm7.74\pm5.90$              &-      &15.6\\ 
       2.1500 &$8.8^{+4.3}_{-3.5}$    &<16.6  &2.8  &0.1016 &1.040  &$113.00^{+55.22}_{-44.95}\pm7.12$  &<213.23&3.1\\ 
       2.1750 &$13.8\pm4.3$           &<22.5  &10.6  &0.1045 &1.040  &$46.02\pm14.34\pm2.90$             &<75.05 &3.4\\ 
       2.2000 &$26.2\pm8.0$           &-      &13.7  &0.1046 &1.040  &$67.40\pm20.58\pm4.25$             &-      &4.8\\        
       2.2324 &$8.4^{+5.0}_{-4.3}$    &<16.5  &11.9  &0.1046 &1.041  &$24.84^{+14.79}_{-12.72}\pm1.56$   &<48.80 &2.2\\ 
       2.3094 &$0.4^{+4.1}_{-3.3}$    &<8.4   &21.1  &0.1036 &1.041  &$0.67^{+6.91}_{-5.56}\pm0.04$      &<14.15 &0.1\\
       2.3864 &$0.0^{+4.4}_{-3.6}$    &<8.4   &22.5  &0.1063 &1.041  &$0.03^{+6.82}_{-5.58}\pm0.00$      &<13.02 &0.0\\
       2.3960 &$3.6^{+6.6}_{-5.8}$    &<16.5  &66.9  &0.1066 &1.041  &$1.85^{+3.40}_{-2.99}\pm0.12$      &<8.50  &0.8\\
       2.5000 &$0.4^{+1.6}_{-0.0}$    &<4.7   &1.1   &0.1127 &1.041  &$11.84^{+47.36}_{-79.91}\pm0.85$   &<139.11&0.1\\
       2.6444 &$9.5^{+4.5}_{-3.8}$    &<16.4  &33.7  &0.1146 &1.039  &$9.08^{+4.30}_{-3.63}\pm0.57$      &<15.67 &3.0\\
       2.6464 &$1.2^{+3.0}_{-2.3}$    &<6.9   &34.0  &0.1151 &1.039  &$1.14^{+2.86}_{-2.19}\pm0.07$      &<6.57  &0.5\\
       2.7000 &$0.0^{+0.6}_{-0.0}$    &<3.9   &1.0   &0.1188 &1.039  &$0.00^{+18.63}_{-0.00}\pm0.00$     &<121.12&0.0\\
       2.8000 &$0.0^{+0.5}_{-0.0}$    &<2.5   &1.0   &0.1185 &1.037  &$0.00^{+15.53}_{-0.00}\pm0.00$     &<77.64 &0.0\\
       2.9000 &$1.5^{+3.8}_{-2.9}$    &<8.7   &105.3   &0.1238 &1.033  &$0.43^{+1.10}_{-0.84}\pm0.03$      &<2.51  &0.5\\
       2.9500 &$1.6^{+2.3}_{-1.6}$    &<6.5   &15.9  &0.1208 &1.029  &$3.08^{+4.43}_{-3.08}\pm0.19$      &<12.51 &1.6\\
       2.9810 &$3.4^{+2.9}_{-2.1}$    &<8.5   &16.1  &0.1266 &1.025  &$6.30^{+5.38}_{-3.89}\pm0.40$      &<15.76 &1.9\\
       3.0000 &$0.0^{+1.1}_{-2.1}$    &<3.2   &15.9  &0.1270 &1.021  &$0.00^{+2.07}_{-3.95}\pm0.00$      &<6.02  &0.0\\
       3.0200 &$0.3^{+1.5}_{-0.8}$    &<3.2   &17.3  &0.1262 &1.014  &$0.52^{+2.60}_{-1.39}\pm0.03$      &<5.55  &0.3\\
       3.0800 &$0.9^{+3.0}_{-2.1}$    &<6.5   &126.2   &0.1298 &0.915  &$0.24^{+0.79}_{-0.55}\pm0.02$      &<1.70  &0.4\\
       \hline
\end{tabular}
}
\caption{The Born cross sections for $e^{+}e^{-}\to\omega\eta^{\prime}$. All definitions of symbols are the same as those in eq.~(\ref{cs}). Upper limits are given at the 90\% confidence level. For the Born cross section $\sigma$, the first uncertainty is statistical and the second is systematic. The VP column lists the vacuum polarization correction factor.}
\label{tab:crosssection}
\end{table}

Various sources of systematic uncertainties concerning the measurement of the Born cross sections are investigated; these include the integrated luminosity, the charged track efficiency, the photon reconstruction efficiency, the PID efficiency, the kinematic fit, the requirement on $M(\pi^{+}\pi^{-}\pi^{0})$, ISR and VP correction factors, input branching fractions, and the fit to the $M(\gamma\pi^{+}\pi^{-})$ distribution. These sources are described as follows.

\begin{itemize}
    \item The uncertainty associated with the integrated luminosity is $1\%$, which is estimated by using large angle Bhabha events in ref.~\cite{besiii_lumin}.
    \item The uncertainties related to the tracking and PID efficiencies of charged pions are investigated using a control sample of $e^{+}e^{-}\to K^{+}K^{-}\pi^{+}\pi^{-}$~\cite{besiii_kk,bes_kskl}. The uncertainties on the tracking and PID efficiencies are estimated to be 1\% per charged pion track.
    \item The uncertainty concerning the photon detection efficiency is studied with a control sample of $e^{+}e^{-}\to K^{+}K^{-}\pi^{+}\pi^{-}\pi^{0}$~\cite{besiii_pi0}. The result shows that the difference in detection efficiency between data and MC simulation is 1\% per photon.
    \item The uncertainties arising from the $\chi^{2}_{\rm{4C}}$ requirement in the kinematic fit and the $\omega$ invariant mass requirement are evaluated from a control sample $J/\psi\to\omega\eta^{\prime}$. The averaged difference between the data and MC simulation of $J/\psi\to\omega\eta^{\prime}$ by using each requirement or not is taken as the systematic uncertainty.
    \item The uncertainty of the VP and ISR correction factors is obtained with the accuracy of the radiation function, which is about 0.5\%~\cite{VP}, and has an additional contribution from the cross section line shape, which is estimated by varying the model parameters of the fit to the cross section. Considering correlations among the parameters, all parameters are randomly generated using a correlated multi-variable Gaussian function and the resulting parametrization of the line shape is used to recalculate $(1+\delta)\epsilon$ and the corresponding cross section. This procedure is repeated one thousand times and the standard deviation of the resulting cross section is considered. The systematic uncertainty associated with the VP and ISR correction factors is assigned as their quadratic sum~\cite{besiii_kk}.
    \item The uncertainty associated with the quoted branching fractions from the PDG~\cite{PDG} is $1.5\%$, including the effects of $\mathcal{B}_{\omega\to\pi^{+}\pi^{-}\pi^{0}}=(89.2\pm0.7)\%$, $\mathcal{B}_{\pi^{0}\to\gamma\gamma}=(98.823\pm0.034)\%$ and $\mathcal{B}_{\eta^{\prime}\to\gamma\pi^{+}\pi^{-}}=(29.5\pm0.4)\%$.
    \item The uncertainty caused by the fit to the $M(\gamma\pi^{+}\pi^{-})$ distribution includes the descriptions of the signal shape, background shape and fit range and is estimated by the control sample $J/\psi\to\omega\eta^{\prime}$. The nominal MC-simulated shape convolved with a Gaussian function is replaced by a pure MC-simulated shape, and the difference is taken as the uncertainty. Replacing the nominal background shape by a first-order Chebychev function, the deviation from the nominal result is taken as the uncertainty. The fit range is varied by $\pm10$ MeV$/c^{2}$ at both boundaries, and the largest difference is taken as the uncertainty. The uncertainties from these three sources are added in quadrature and taken as the total uncertainty from the $M(\gamma\pi^{+}\pi^{-})$ fit. 
\end{itemize}

    Table~\ref{tab:uncertainty} summarizes all the systematic uncertainties related to the Born cross section for each energy point, where the sources of the uncertainties tagged with `*' are assumed to be 100\% correlated among c.m.\ energies. The total systematic uncertainty for each energy point is calculated as the quadratic sum of the individual uncertainties, assuming they are independent.

\begin{table}[tb]
    \centering
    \begin{tabular}{c|cccccccccc}
    \hline
    $\sqrt{s}$~(GeV) &$\mathcal{L}^{*}$  &$\rm{Track}^{*}$ &$\rm{Pho.}^{*}$ &$\rm{PID}^{*}$ &$\rm{Kin.}^{*}$ &$\omega~\rm{Mass}^{*}$ &$\rm{Rad.}$ &$\rm{Br}^{*}$ &Fit$^{*}$ &Sum \\ \hline
    2.0000 &1.0  &4.0  &3.0  &4.0 &0.2 &0.9 &0.6 &1.5 &2.4 &7.1\\ 
    2.0500 &1.0  &4.0  &3.0  &4.0 &0.2 &0.9 &0.7 &1.5 &2.4 &7.2\\ 
    2.1000 &1.0  &4.0  &3.0  &4.0 &0.2 &0.9 &1.5 &1.5 &2.4 &7.3\\ 
    2.1250 &1.0  &4.0  &3.0  &4.0 &0.2 &0.9 &1.5 &1.5 &2.4 &7.3\\ 
    2.1500 &1.0  &4.0  &3.0  &4.0 &0.2 &0.9 &1.1 &1.5 &2.4 &7.2\\ 
    2.1750 &1.0  &4.0  &3.0  &4.0 &0.2 &0.9 &0.8 &1.5 &2.4 &7.2\\ 
    2.2000 &1.0  &4.0  &3.0  &4.0 &0.2 &0.9 &0.7 &1.5 &2.4 &7.2\\        
    2.2324 &1.0  &4.0  &3.0  &4.0 &0.2 &0.9 &0.6 &1.5 &2.4 &7.2\\ 
    2.3094 &1.0  &4.0  &3.0  &4.0 &0.2 &0.9 &0.6 &1.5 &2.4 &7.2\\
    2.3864 &1.0  &4.0  &3.0  &4.0 &0.2 &0.9 &0.6 &1.5 &2.4 &7.2\\
    2.3960 &1.0  &4.0  &3.0  &4.0 &0.2 &0.9 &0.6 &1.5 &2.4 &7.2\\
    2.5000 &1.0  &4.0  &3.0  &4.0 &0.2 &0.9 &0.6 &1.5 &2.4 &7.2\\
    2.6444 &1.0  &4.0  &3.0  &4.0 &0.2 &0.9 &0.5 &1.5 &2.4 &7.2\\
    2.6464 &1.0  &4.0  &3.0  &4.0 &0.2 &0.9 &0.6 &1.5 &2.4 &7.2\\
    2.7000 &1.0  &4.0  &3.0  &4.0 &0.2 &0.9 &0.6 &1.5 &2.4 &7.2\\
    2.8000 &1.0  &4.0  &3.0  &4.0 &0.2 &0.9 &0.5 &1.5 &2.4 &7.2\\
    2.9000 &1.0  &4.0  &3.0  &4.0 &0.2 &0.9 &0.5 &1.5 &2.4 &7.2\\
    2.9500 &1.0  &4.0  &3.0  &4.0 &0.2 &0.9 &0.5 &1.5 &2.4 &7.2\\
    2.9810 &1.0  &4.0  &3.0  &4.0 &0.2 &0.9 &0.5 &1.5 &2.4 &7.2\\
    3.0000 &1.0  &4.0  &3.0  &4.0 &0.2 &0.9 &0.5 &1.5 &2.4 &7.2\\
    3.0200 &1.0  &4.0  &3.0  &4.0 &0.2 &0.9 &0.5 &1.5 &2.4 &7.2\\
    3.0800 &1.0  &4.0  &3.0  &4.0 &0.2 &0.9 &0.5 &1.5 &2.4 &7.2\\
    \hline
    \end{tabular}
    \caption{Systematic uncertainties~(in \%) in the Born cross section of $e^{+}e^{-}\to\omega\eta^{\prime}$ at each energy point. These represent the uncertainties in the estimated effects of the integrated luminosity~($\mathcal{L}$), tracking efficiency ($\rm{Track}$), photon reconstruction efficiency ($\rm{Pho.}$), PID efficiency ($\rm{PID}$), kinematic fit ($\rm{Kin.}$), $\omega$ mass window ($\omega~\rm{Mass}$), VP and ISR correction factor ($\rm{Rad.}$), branching fraction ($\rm{Br}$) and $M(\gamma\pi^{+}\pi^{-})$ fit ($\rm{Fit}$). The total uncertainty~(Sum) is obtained by summing the individual contributions in quadrature.}
    \label{tab:uncertainty}
\end{table}
    \section{Fit to the line shape}
\begin{figure}[t]
        \centering
        \begin{overpic}[width=0.45\textwidth]{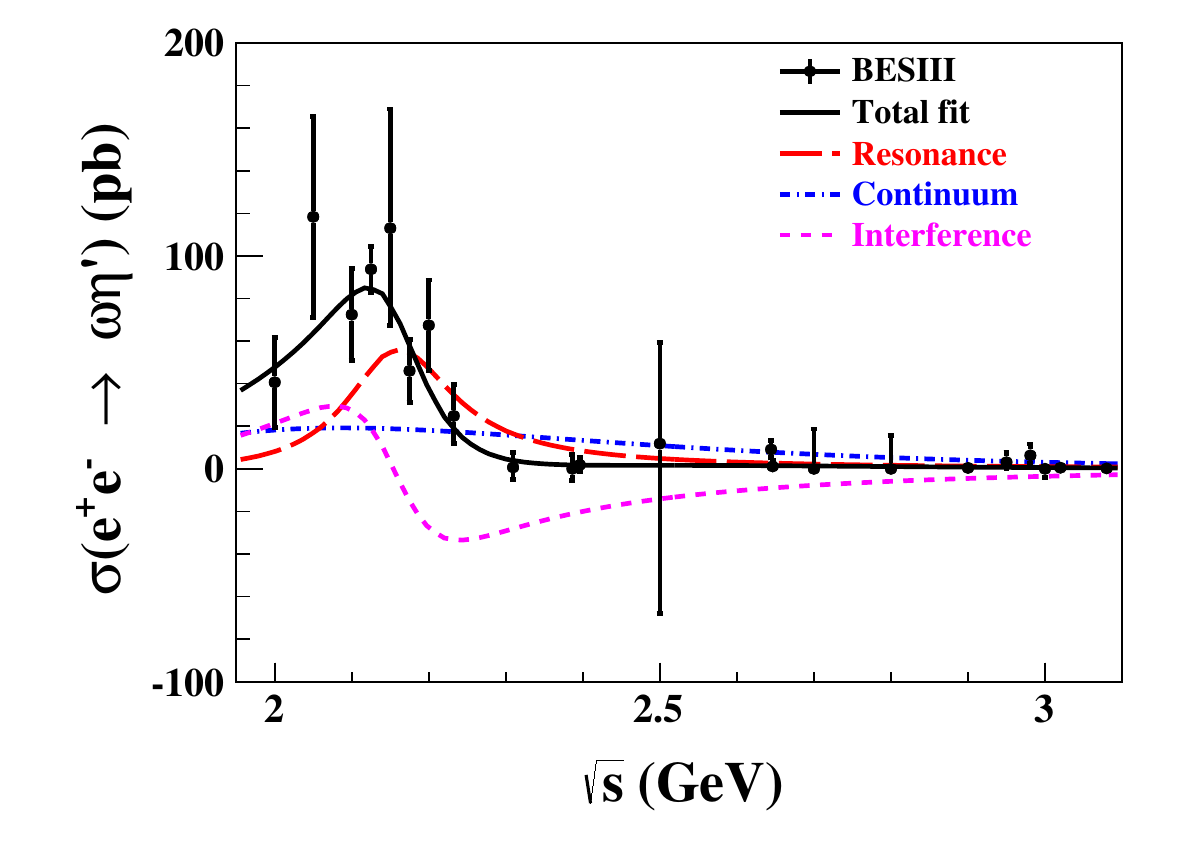}
        \put(87,63){(a)}
        \end{overpic}
        \begin{overpic}[width=0.45\textwidth]{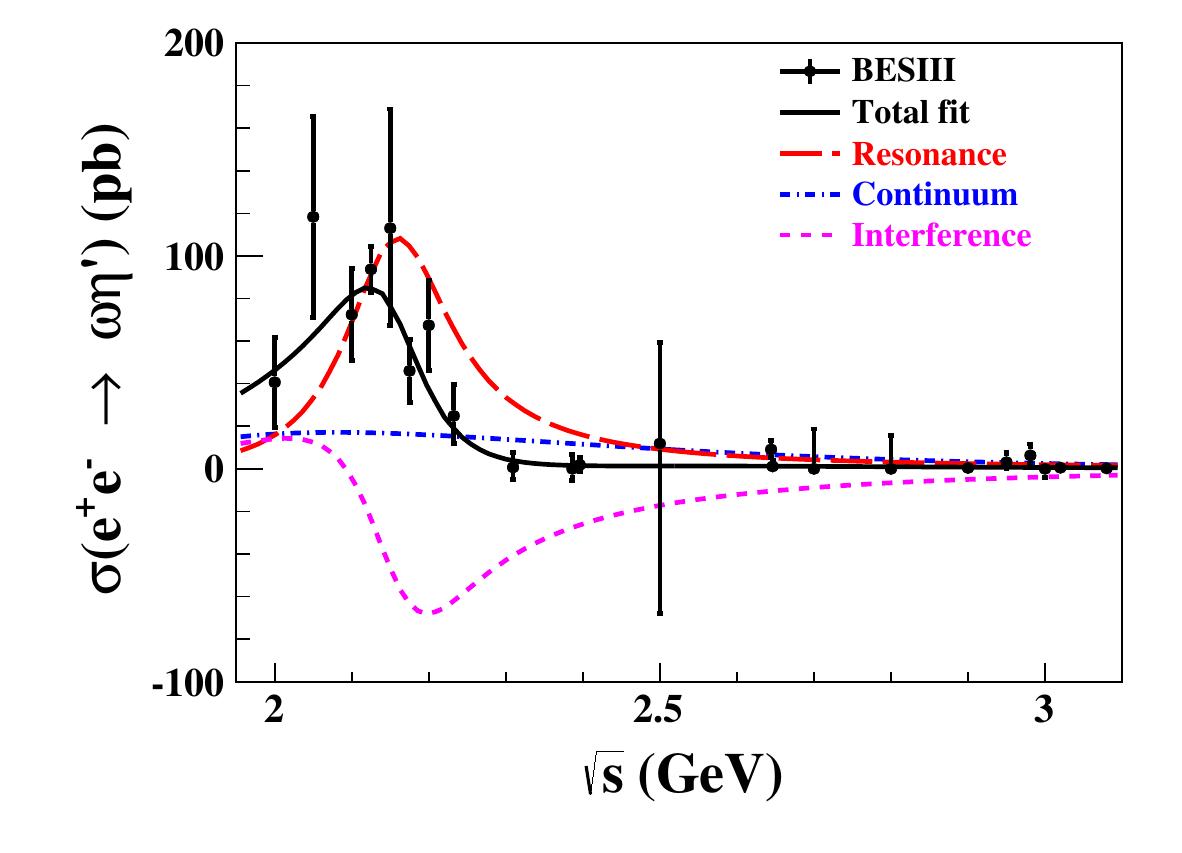}
        \put(87,63){(b)}
        \end{overpic}
        \flushleft
        \caption{The Born cross section and fit curves, (a) and (b), corresponding to the two solutions shown in table~\ref{tab:fitresult}. Dots with error bars are BESIII data, where errors include both statistical and systematic uncertainties. The solid~(black) curves represent the total fit result, the dashed~(red) curves are for the intermediate state and the dash-dotted~(blue) curves are for the continuum component, and the intensive dash-dotted~(magenta) curves are for the interference between the resonance and continuum components.}
        \label{fig:fitresult}
    \end{figure}

    The measured Born cross sections are shown in figure~\ref{fig:fitresult}, where a clear structure is visible around 2.2 GeV. A $\chi^{2}$ fit, incorporating the correlated and uncorrelated uncertainties among different energy points, is performed to extract resonance parameters. The fit probability density function is a coherent sum of a continuum component $f_{1}$ and a resonant component $f_{2}$. The cross section is modeled as
    \begin{equation}
        \sigma=\frac{12\pi}{s^{3/2}}|f_{1}+e^{i\phi}f_{2}|^{2}PS(\sqrt{s}),
    \end{equation}
    where $\phi$ is the relative phase between these two components, and $PS(\sqrt{s})$ is the phase-space factor~\cite{bes_omegaeta} given by $PS(\sqrt{s})=q^{3}$, where $q$ is the $\omega$ momentum in the $e^{+}e^{-}$ c.m.\ frame calculated for the mass value $M_{\omega}=0.78265$ GeV/$c^{2}$~\cite{PDG}. The contribution from lower-mass vector states is considered in the description of $f_{1}$ as
    \begin{equation}
        f_{1}=C_{0}\cdot e^{-p_{0}(\sqrt{s}-M_{\rm{th}})},
    \end{equation}
    where $C_{0}$ and $p_{0}$ are free parameters, and $M_{\rm{th}}=1.7404$ GeV/$c^{2}$ is the mass threshold for $\omega\eta^{\prime}$ production~\cite{PDG,besiii_kkpi}. The resonant amplitude $f_{2}$ is described by a Breit-Wigner~(BW) function:
    \begin{equation}
        f_{2}=\sqrt{\frac{\Gamma_{R}^{e^{+}e^{-}}\cdot B_{R}^{\omega\eta^{\prime}}}{PS(m_{R})}}\frac{m_{R}^{3/2}\sqrt{\Gamma_{R}}}{s-m_{R}^{2}+i\sqrt{s}\Gamma_{R}},
    \end{equation}
    where $m_{R}$ and $\Gamma_{R}$ are the mass and width of the resonant structure, $\Gamma_{R}^{e^{+}e^{-}}$ is its partial width to $e^{+}e^{-}$, and $B_{R}$ is the branching fraction of $R\to\omega\eta^{\prime}$.

    In total, there are six free parameters in the fit: $\phi,C_{0},p_{0},m_{R},\Gamma_{R}$ and the product of $\Gamma_{R}^{e^{+}e^{-}}B_{R}$. The fit results are shown in figure~\ref{fig:fitresult}, and resonance parameters are listed in table~\ref{tab:fitresult}. The fit has two solutions with identical mass and width for the resonance. The two solutions for the phase $\phi$ and $\Gamma_{R}^{e^{+}e^{-}}B_{R}$ are consistent with each other considering their uncertainties. The fit quality is $\chi^{2}/\rm{ndf}=13.5/16$, where ndf is the number of degrees of freedom. The mass and width of the resonance are $M_R=(2153\pm30)~\rm{MeV}/c^{2}$ and $\Gamma_{R}=(167\pm77)~\rm{MeV}$, respectively, where the uncertainties are statistical only. The significance of the resonance is determined to be $9.6\sigma$ by comparing the change of the $\chi^{2}$ in fits with and without the resonance and considering the change in ndf. The uncertainties~(statistical and systematic) of the measured Born cross sections have been considered when fitting the line shape, extracting the resonance parameters and estimating the significance of the resonance.

    \begin{table}[t]
    \centering
    \begin{tabular}{c|cc}  
    \hline
         Parameter      &Solution 1 &Solution 2\\\hline

         $M_{R}$ (MeV/$c^2$) &\multicolumn{2}{c}{$2153\pm30(\rm{stat.})\pm31(\rm{syst.})$} \\ 
         $\Gamma_{R}$ (MeV)  &\multicolumn{2}{c}{$167\pm77(\rm{stat.})\pm7(\rm{syst.})$}  \\ 
         $\phi$ (rad)        &$3.78\pm0.24(\rm{stat.})\pm0.12(\rm{syst.})$ &$3.16\pm0.4(\rm{stat.})\pm0.2(\rm{syst.})$  \\
         $\Gamma^{e^+e^-}_{R}\cal{B}_{R}^{\omega\eta'}$ (eV) &$5.72\pm1.68(\rm{stat.})\pm1.5(\rm{syst.})$ &$2.99\pm1.68(\rm{stat.})\pm1.2(\rm{syst.})$\\
         Significance       &\multicolumn{2}{c}{$9.6\sigma$}  \\
        \hline
    \end{tabular}
    \caption{The obtained resonance parameters.}
    \label{tab:fitresult}
    \end{table}

    The systematic uncertainties of the resonant parameters are due to the c.m.\ energy from BEPCII and the formula used in the fit procedure. The uncertainty of the c.m.\ energy calibration is estimated as $0.1\%$. It is ignored in the determination of the resonance parameters~\cite{besiii_lumin}. To evaluate the systematic uncertainty associated with the fit formula, the continuum term $C_{0}\cdot e^{-p_{0}(\sqrt{s}-M_{\rm{th}})}$ is replaced with an exponential function of the form $c_1/s^{c_2}$, where $c_1$ and $c_{2}$ are free parameters and the width $\Gamma_{R}$ is replaced with an energy-dependent width of the form $\Gamma_{R}(\sqrt{s})=\Gamma_{R}\cdot\frac{PS(\sqrt{s})}{PS(M_{R})}$~\cite{besiii_kkpi}. The deviation of the obtained parameters from the nominal results are taken as the systematic uncertainties. The resonance parameters, which have been considered systematically, are listed in table~\ref{tab:fitresult}.
    
    \section{Summary}
    In summary, the Born cross sections of the process $e^{+}e^{-}\to\omega\eta^{\prime}$ are measured at center-of-mass energies from 2.000 to 3.080 GeV using 22 data samples corresponding to a total integrated luminosity of 650 $\rm{pb}^{-1}$ collected by the BESIII detector. By analyzing the energy dependence of the cross sections, a structure is observed with a statistical significance of 9.6$\sigma$ with mass $M_R=(2153\pm30\pm31)~{\rm{MeV}}/c^{2}$ and width $\Gamma_{R}=(167\pm77\pm7)~\rm{MeV}$, where the first and second uncertainties are statistical and systematic, respectively.

    Compared to the resonance parameters for structures around 2.2 GeV obtained from the previous BESIII measurements via $e^{+}e^{-}\to\omega\eta$~\cite{bes_omegaeta}, $e^{+}e^{-}\to\omega\pi^{0}\pi^{0}$~\cite{bes_omegapi0pi0} and $e^{+}e^{-}\to\omega\pi^{+}\pi^{-}$~\cite{bes_omegapippim}, $M_{1}=2176\pm24$ ${\rm{MeV}}/c^{2}$, $\Gamma_{1}=89\pm50$ $\rm{MeV}$ and \mbox{$M_{2}=2232\pm19$ ${\rm{MeV}}/c^{2}$}, \mbox{$\Gamma_{2}=93\pm53$ $\rm{MeV}$}, respectively, the parameters obtained in this work are in agreement with them within $3\sigma$. 

    Supposing the resonances around 2.2 GeV observed in $R\to\omega\eta$ and $R\to\omega\eta^{\prime}$ are the same, and $R$ only contains $u\bar{u}$ and $d\bar{d}$ quarks, the ratio between branching fractions of $R\to\omega\eta$ and $R\to\omega\eta^{\prime}$ can be written as 
    \begin{equation}
        \frac{B_{R}^{\omega\eta^{\prime}}}{B_{R}^{\omega\eta}}=\left| \frac{\cos(\Theta_{0}-\Theta)}{\sin(\Theta_{0}-\Theta)}\right|^{2}\frac{\Omega_{\eta^{\prime}}}{\Omega_{\eta}},
    \end{equation}
    where $\Theta_{0}=35.3^\circ$ is the ideal mixing angle between pure $s\bar{s}$ and $u\bar{u}-d\bar{d}$, and $\Omega_{\eta^{\prime}}(\Omega_{\eta})$ is the phase-space factor including the P-wave effect in the decay~\cite{Zhukai,Zhukai2}. Based on the suggested mixing angle $\Theta=-14.1\pm2.8^{\circ}$ calculated by lattice QCD~\cite{PDG,LQCD_mix}, the ratio between $\Gamma^{e^+e^-}_{R}\cal{B}_{R}^{\omega\eta^{\prime}}$ and $\Gamma^{e^+e^-}_{R}\cal{B}_{R}^{\omega\eta}$ is $0.34\pm0.07$. Combining our results with the previous BESIII measurements of the $e^{+}e^{-}\to\omega\eta$ process~\cite{bes_omegaeta}, and taking into account the mutiple solutions for the resonance parameters, these ratios are measured to be $13.30\pm6.06\pm3.61$, \mbox{$2.39\pm1.63\pm1.01$}, \mbox{$4.58\pm2.21\pm1.34$} and \mbox{$6.95\pm4.60\pm2.83$}, where the first and second uncertainties are statistical and systematic, respectively. All of them are greater than $0.34\pm0.07$ based on the $\eta-\eta^{\prime}$ mixing angle given by lattice QCD. Future experimental and theoretical studies will be helpful to improve the knowledge of excited $\omega$ states around 2.2 GeV.

    \acknowledgments
    The BESIII Collaboration thanks the staff of BEPCII, the IHEP computing center and the supercomputing center of USTC for their strong support. This work is supported in part by National Key R\&D Program of China under Contracts Nos. 2020YFA0406400, 2020YFA0406300; National Natural Science Foundation of China (NSFC) under Contracts Nos. 11635010, 11735014, 11835012, 11935015, 11935016, 11935018, 11961141012, 12022510, 12025502, 12035009, 12035013, 12192260, 12192261, 12192262, 12192263, 12192264, 12192265, 12275320, 11625523, 11705192, 11950410506, 12061131003, 12105276, 12122509; the Chinese Academy of Sciences (CAS) Large-Scale Scientific Facility Program; the CAS Center for Excellence in Particle Physics (CCEPP); Joint Large-Scale Scientific Facility Funds of the NSFC and CAS under Contracts Nos. U1832207, U1732263, U1832103, U2032111; CAS Key Research Program of Frontier Sciences under Contracts Nos. QYZDJ-SSW-SLH003, QYZDJ-SSW-SLH040; 100 Talents Program of CAS; The Institute of Nuclear and Particle Physics (INPAC) and Shanghai Key Laboratory for Particle Physics and Cosmology; ERC under Contract No. 758462; European Union's Horizon 2020 research and innovation programme under Marie Sklodowska-Curie grant agreement under Contract No. 894790; German Research Foundation DFG under Contracts Nos. 443159800, 455635585, Collaborative Research Center CRC 1044, FOR5327, GRK 2149; Istituto Nazionale di Fisica Nucleare, Italy; Ministry of Development of Turkey under Contract No. DPT2006K-120470; National Research Foundation of Korea under Contract No. NRF-2022R1A2C1092335; National Science and Technology fund; National Science Research and Innovation Fund (NSRF) via the Program Management Unit for Human Resources \& Institutional Development, Research and Innovation under Contract No. B16F640076; Polish National Science Centre under Contract No. 2019/35/O/ST2/02907; The Royal Society, UK under Contracts Nos. DH140054, DH160214; The Swedish Research Council; U. S. Department of Energy under Contract No. DE-FG02-05ER41374.

\providecommand{\href}[2]{#2}\begingroup\raggedright\endgroup
\newpage
\section*{The BESIII collaboration}
\addcontentsline{toc}{section}{The BESIII collaboration}
M.~Ablikim$^{1}$, M.~N.~Achasov$^{5,b}$, P.~Adlarson$^{74}$, X.~C.~Ai$^{80}$, R.~Aliberti$^{35}$, A.~Amoroso$^{73A,73C}$, M.~R.~An$^{39}$, Q.~An$^{70,57}$, Y.~Bai$^{56}$, O.~Bakina$^{36}$
% ...
\\
\vspace{0.2cm}
(BESIII Collaboration)\\
\vspace{0.2cm} {\it
$^{1}$ Institute of High Energy Physics, Beijing 100049, People's Republic of China\\
% ...
$^{5}$ Budker Institute of Nuclear Physics SB RAS (BINP), Novosibirsk 630090, Russia\\
% ...
$^{35}$ Johannes Gutenberg University of Mainz, Johann-Joachim-Becher-Weg 45, D-55099 Mainz, Germany\\
$^{36}$ Joint Institute for Nuclear Research, 141980 Dubna, Moscow region, Russia\\
% ...
$^{39}$ Liaoning Normal University, Dalian 116029, People's Republic of China\\
% ...
$^{56}$ Southeast University, Nanjing 211100, People's Republic of China\\
$^{57}$ State Key Laboratory of Particle Detection and Electronics, Beijing 100049, Hefei 230026, People's Republic of China\\
% ...
$^{70}$ University of Science and Technology of China, Hefei 230026, People's Republic of China\\
% ...
$^{73}$ University of Turin and INFN, (A)University of Turin, I-10125, Turin, Italy; (B)University of Eastern Piedmont, I-15121, Alessandria, Italy; (C)INFN, I-10125, Turin, Italy\\
$^{74}$ Uppsala University, Box 516, SE-75120 Uppsala, Sweden\\
% ...
$^{80}$ Zhengzhou University, Zhengzhou 450001, People's Republic of China\\

\vspace{0.2cm}
\noindent
% ...
$^{b}$ Also at the Novosibirsk State University, Novosibirsk, 630090, Russia\\
% ...
}
\end{document}